\documentclass[]{llncs}
\usepackage[T1]{fontenc}
\usepackage{multirow}
\usepackage{enumitem}
\usepackage{graphicx}
\usepackage{amsmath}
\begin{document}
\title{Towards an In-Depth Comprehension of Case Relevance for Better Legal Retrieval}

%
%
\author{Haitao Li\inst{1,2}\orcidID{0009-0006-8766-8610} \and
You Chen\inst{1,2}\orcidID{0009-0008-9873-4315} \and
Zhekai Ge\inst{3}\orcidID{0009-0005-5619-8591} \and
Qingyao Ai \inst{1,2}\orcidID{0000-0002-5030-709X} \and
Yiqun Liu \inst{1,2}\orcidID{0000-0002-0140-4512} \and
Quan Zhou \inst{4,2} \orcidID{0009-0003-8097-4621} \and
Shuai Huo \inst{4,2} \orcidID{0009-0007-1276-0268}
}
\authorrunning{Li. Author et al.}
%
\institute{DCST, Tsinghua University, China \and
Quan Cheng Laboratory, China \and
Columbia University, USA \and
MegaTech.AI, China \\
\email{liht22@mails.tsinghua.edu.cn}
}
\maketitle              
\begin{abstract}
Legal retrieval techniques play an important role in preserving the fairness and equality of the judicial system. As an annually well-known international competition, COLIEE aims to advance the development of state-of-the-art retrieval models for legal texts. 
This paper elaborates on the methodology employed by the TQM team in COLIEE2024.
Specifically, we explored various lexical matching and semantic retrieval models, with a focus on enhancing the understanding of case relevance.
Additionally, we endeavor to integrate various features using the learning-to-rank technique.
Furthermore, fine heuristic pre-processing and post-processing methods have been proposed to mitigate irrelevant information.
Consequently, our methodology achieved remarkable performance in COLIEE2024, securing first place in Task 1 and third place in Task 3.
We anticipate that our proposed approach can contribute valuable insights to the advancement of legal retrieval technology.

\end{abstract}
\section{Introduction}
Efficient legal retrieval is essential in the judicial process. It supports lawyers in argumentation, guides judges in decision-making, and aids scholars in analyzing legal trends.  With the evolution of the legal field into the digital age, the ability to efficiently navigate vast legal databases with advanced search techniques is essential for the maintenance of justice ensuring the judicial fairness~\cite{shao2020bert,bench2012history,yu2022explainable,althammer2021dossier,li2023thuircoliee,li2023thuircoliee2,li2023sailer}.

The Competition on Legal Information Extraction/Entailment (COLIEE) has emerged as a significant platform for advancing the state-of-the-art in legal information processing and retrieval. The competition consists of several tasks focusing on two categories: legal retrieval and legal entailment.

This year, our team TQM primarily focused on participating in the legal retrieval tasks, i.e. Task 1 and Task 3. Task 1 involves retrieving relevant documents to support a given query case within the case law system. Task 3 involves retrieving civil law related to Japanese Legal Bar exam questions under the statutory law system. Through a thorough comprehension of case relevance, the TQM team achieved commendable results in COLIEE2024.

In legal practice, case relevance is complex and differs from that of conventional web search~\cite{ma2021lecard,shao,li2023lecardv2}. In the context of legal retrieval, relevance transcends mere lexical matches or semantic similarities. The relevance of legal cases usually involves an in-depth analysis of the facts of the case, legal principles, and prior jurisprudence~\cite{li2023lecardv2,shao2023understanding,li2024blade}. This requires the retrieval system to understand not only the words and concepts in the text, but also to gain insight into their interactions within a particular legal framework. Traditional methods often prove inadequate in capturing the nuanced aspects that determine case relevance, including the construction of legal arguments, key legal facts, and the particular nature of applicable laws.

Therefore, during COLIEE2024, our team, TQM, not only investigated the effectiveness of established methods in legal retrieval but also explored new strategies to improve the model's understanding of case relevance.
Specifically, within the traditional lexical matching approach, we employed BM25\_ngram to underscore the significance of law-specific terms in determining relevance. Additionally, in the semantic similarity approach, we utilized the translation process between different structures of legal cases to deepen the understanding of key facts. 
Subsequently, we employed learning-to-rank techniques to integrate different features. In addition, we design delicate heuristic pre-processing and post-processing methods to mitigate the impact of irrelevant information. 
In conclusion, the official results reveal our team's remarkable achievement, attaining first place in Task 1 and third place in Task 3. This shows the effectiveness of our design approach.

The paper is structured as follows: Section 2 offers an overview of foundational concepts in legal case retrieval and dense retrieval. Section 3 elaborates on the COLIEE2024 legal case retrieval task, encompassing its description, datasets, and evaluation metrics. Section 4 delves into the technical aspects of the study. Following this, Section 5 presents the results of our experiments. The paper concludes with Section 6, summarizing key findings and outlining directions for future research.

\section{Related Work}
\subsection{Legal Retrieval}
In the area of legal retrieval, the integration of deep learning techniques has become foundational, giving rise to a plethora of methodologies such as CNN-based models~\cite{tran2019building}, BiDAF~\cite{seo2016bidirectional}, and SMASH-RNN~\cite{jiang2019semantic}, among others. 
Generative transformers have emerged as the preferred architecture in this domain, notably powering innovations like LEGAL-BERT~\cite{chalkidis2020legal} and Lawformer~\cite{xiao2021lawformer}.
Besides, Jiang et al.~\cite{jiang2020cross} demonstrated improvements in cross-lingual retrieval, by using Multilingual BERT to handle the linguistic space in legal documentation. Recent contributions further enriched this field. By focusing on context-aware citation recommendations~\cite{huang2021context} and graph-based legal reasoning~\cite{zhang2023cfgl}, we can significantly enhance relevance and semantic richness of case retrieval methods. Also, 
Li et al. proposed SAILER~\cite{li2023sailer}, which utilizes the structure of legal documents for pre-training and achieves the best results on some legal benchmarks.
These developments highlights the potential of transformative strides in AI and machine learning to legal information retrieval.

\subsection{Dense Retrieval}
A radical departure from traditional retrieval has emerged through dense retrieval, which leverages dual encoders to map the queries and documents into dense embeddings and capture intricate contextual nuances~\cite{xie2023t2ranking,dong2023i3}. This method has been progressively improved through a series of innovative works: Zhan et al.\cite{li2023constructing} introduced dynamic negative sampling to refine the matching process and Chen et al.\cite{chen2022axiomatically} unveiled ARES that incorporates retrieval axioms during pre-training, which substantially improved performance. Similarly, Karpukhin et al. \cite{karpukhin2020dense} introduced DPR (Dense Passage Retrieval) which surpassed traditional IR methods by a large margin in large-scale open-domain question-answering tasks, and Xiong et al.\cite{xiong2020approximate}introduced ANCE (Approximate Nearest Neighbor Negative Contrastive Learning), which dynamically updated the negative samples and further optimized the retrieval process. These studies, demonstrate the potential of dense retrieval to revolutionize IR technologies, providing more accurate results across various applications.

\section{Task Overview}
\subsection{Task1.The Case Law Retrieval Task}
\subsubsection{Task Description}

The Competition on Legal Information Extraction/Entailment (COLIEE), an annual international contest, is committed to advancing state-of-the-art methodologies in legal text processing. 
In COLIEE2024, four tasks are presented, with our exclusive focus directed towards the legal retrieval task.

Task 1, referred to as the Case Law Retrieval task, involves the identification of supporting cases that substantiate the decisions of query cases within an extensive corpus. Formally, for a given query case denoted as $q$ and a set of candidate cases represented by $S$, the objective is to identify all supporting cases, designated as $S_q^* = \{S_1, S_2, ..., S_n\}$ from the extensive candidate pool. Participants are allowed to submit any number of supporting cases for each individual query in this task. Hence, it is also crucial to identify the conditions fulfilled by the relevant cases.

The data corpus utilized for Task 1 comprises a collection of case law documents from the Federal Court of Canada, provided by Compass Law. Detailed statistics of this dataset are presented in Table \ref{satistics}. Through our analysis, we find that there is a significant difference in the average number of relevant documents per query between the COLIEE2023 training and test sets. Therefore, we similarly consider possible bias for effective post-processing in COLIEE2024. We employ the test set of COLIEE2023 as the validation set and and apply the best parameters in COLIEE2023 to COLIEE2024.

\begin{table*}[t]
\centering
\caption{Dataset statistics of COLIEE Task 1.}
\scriptsize
\begin{tabular}{l|ll|ll|ll|ll}
\hline
\multirow{2}{*}{}                        & \multicolumn{2}{l|}{COLIEE2021} & \multicolumn{2}{l|}{COLIEE2022} & \multicolumn{2}{l|}{COLIEE2023} & \multicolumn{2}{l}{COLIEE2024} \\
                                         & Train           & Test          & Train           & Test          & Train           & Test          & Train          & Test          \\ \hline
\# of queries                            & 650             & 250           & 898             & 300           & 959             & 319           & 1278           & 400           \\
\# of candidate case per query           & 4415            & 4415          & 3531            & 1263          & 4400            & 1335          & 5616           & 1734          \\
avg \# of relevant candidates/paragraphs & 5.17            & 3.60          & 4.68            & 4.21          & 4.68            & 2.69          & 4.16           & -             \\ \hline
\end{tabular}
\label{satistics}
\end{table*}

\subsubsection{Metrics}
For COLIEE 2024 Task 1, the evaluation metrics will include precision, recall, and the F1-measure:

\begin{equation}
\text { Precision } = \frac{\# T P}{\# T P+\# F P} 
\end{equation}

\begin{equation}
\text { Recall } = \frac{\# T P}{\# T P+\# F N} 
\end{equation}

\begin{equation}
F-\text { measure } = \frac{2 \times \text { Precision } \times \text { Recall }}{\text { Precision }+ \text { Recall }}
\end{equation}

where $\#TP$ represents the total number of accurately retrieved candidate cases across all queries, $\#FP$ denotes the number of incorrectly retrieved candidate cases for all queries, and 
$\#FN$ signifies the count of overlooked noticed candidate paragraphs in all queries. Notably, the evaluation process employed a micro-average approach, where the evaluation measure is computed based on the collective results of all queries. This differs from a macro-average approach, which calculates the evaluation measure for each query individually before averaging these values.

\begin{figure*}[ht]
\centering
\includegraphics[width=0.9\textwidth]{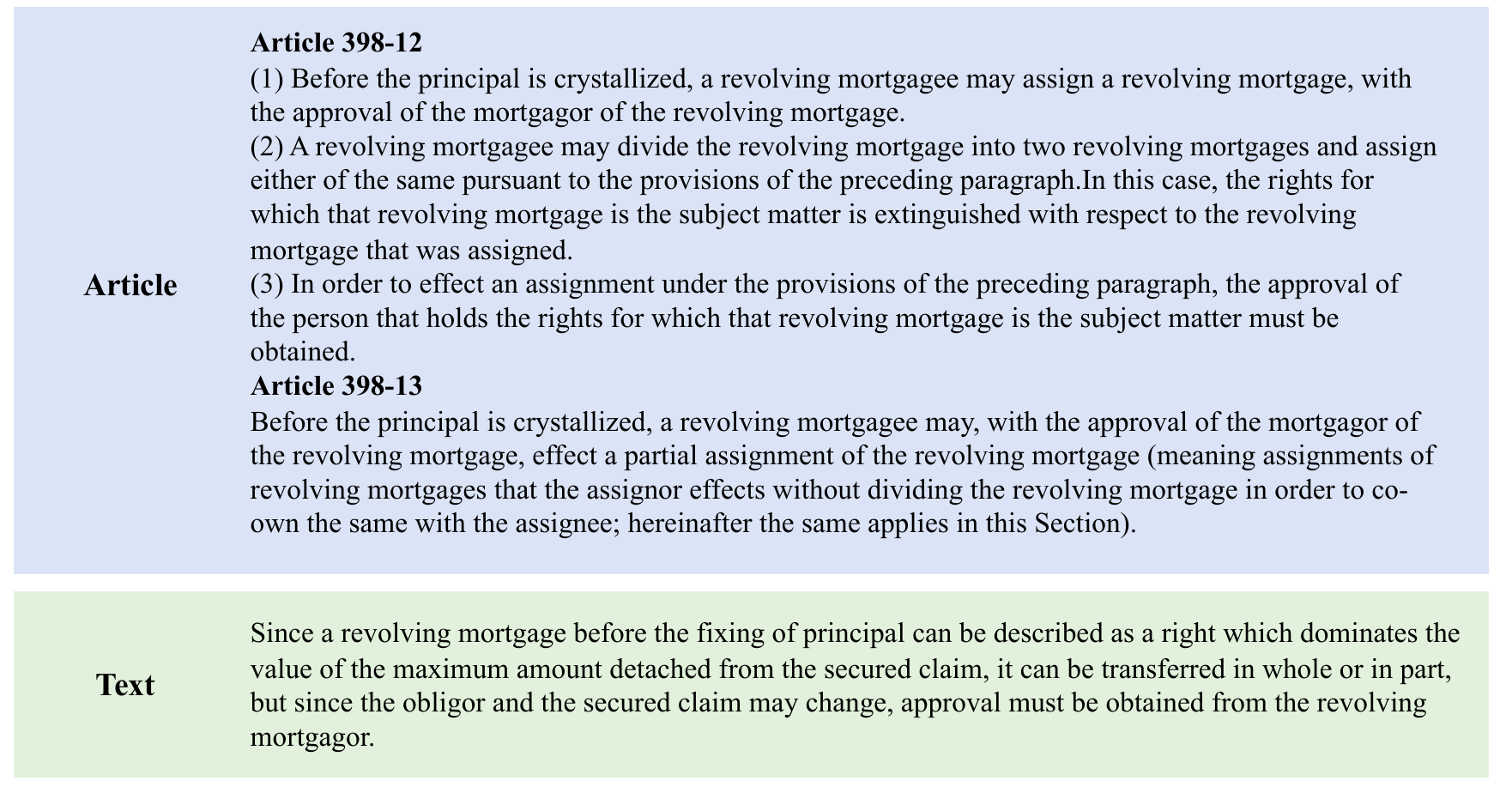}
\vspace{-3mm}
\caption{Example of task3 in COLIEE2024. The label is Yes, which means that the text is relevant to the articles.} 
\label{example}
\end{figure*}

\subsection{Task3.The Statute Law Retrieval Task}
\subsubsection{Task Description}
This task focuses on retrieving civil law articles relevant to a given "Yes/No" question. For a legal bar exam question denoted as $Q$
Q and a set of Japanese Civil Code Articles represented as 
$S = {S_1,...,S_n}$, the objective is to compile a subset 
$E$ from $S$ that aids in answering $Q$. The questions for this task are sourced from Japanese Legal Bar Exams and are translated into English, along with the entire corpus of Japanese Civil Law articles.

The dataset of this task consists of 1097 pairs, a legal corpus (Civil Code) with 768 articles, and 109 test queries.
Participants need to find the relevant articles for the test query. The examples of this dataset are shown in Figure~\ref{example}. More accurately, this task is more like a ranking task, since the candidate set has only 768 legal entries. We selected questions with IDs beginning with R04 with 101 questions to form a validation set. This subset was utilized to conduct evaluations of various models and settings.

\subsubsection{Metrics}
For COLIEE 2024 Task 3, the evaluation criteria include macro-average precision, recall, and F2-measure, diverging from the micro-average measures traditionally used in Task 1.

\begin{equation}
\text { Precision } = \text{Average of}\frac{|\text{Correctly articles for each query|}}{|\text{Retrieved articles for each query}|} 
\end{equation}

\begin{equation}
\text { Recall } = \text{Average of}\frac{|\text{Correctly retrieved articles for each query|}}{|\text{Correct articles for each query}|} 
\end{equation}

\begin{equation}
F-\text { measure } = \frac{5 \times \text { Precision } \times \text { Recall }}{4 \times\text { Precision }+ \text { Recall }}
\end{equation}

\section{Method}
In this section, we present our approach and motivation for the legal case retrieval task in COLIEE2024.

\subsection{Task1.The Case Law Retrieval Task}
In this section, we present our solution in detail for Task 1 of COLIEE2024. Overall, we followed the framework of last year's first place team THUIR~\cite{li2023thuircoliee}. We first pre-process the data to eliminate noisy information. After that, we implemented the classical lexical matching method and the state-of-the-art semantic retrieval model. The difference is that we improve both approaches from the perspective of case relevance. Following this, we use learning to rank to fuse features from different perspectives for better modeling of case relevance. Finally, we propose heuristic post-processing strategies by observing common properties of relevant cases.

\subsubsection{Pre-processing}
Following li et al~\cite{li2023thuircoliee}, we perform the fine data pre-processing before training.
To be specific, our initial step involved the removal of text before the ``[1]'' character in each case document, which typically includes procedural details such as time and court. Subsequently, we eliminated all placeholders, notably ``FRAGMENT\_SUPPRESSED'', to avoid interference in similarity computations. Additionally, in cases where legal documents contained French text, we utilized the Langdetect tool to identify and remove French passages. Documents predominantly in French were translated into English to retain their essential information. In the process of summary extraction, we selectively extracted sections under ``summary'' subheadings, which generally encapsulate key case elements, and integrated these at the beginning of the processed text.
Through preprocessing, Through this pre-processing, we aimed to reduce as much noisy information in the case documents as possible, which does not contribute to the relevance judgment.

\subsubsection{Lexical Matching Models}
In previous competitions, many participants have discovered that traditional lexical matching models can produce competitive results. This phenomenon can be attributed to two primary factors. Firstly, bag-of-words models do not impose limitations on the text length, rendering them well-suited for handling legal case documents with lengthy texts. Secondly, the legal domain encompasses numerous specialized terms, where relevance is often discernible through word matching.
Therefore, in this section, we experimented with the following methods:

\begin{itemize}[leftmargin=*]
    \item[-] \textbf{BM25}~\cite{robertson2009probabilistic} a probabilistic relevance model grounded in the bag-of-words concept, calculates relevance between a query $q$ and a document $d$. The formulation of BM25 is presented as follows:
    \begin{equation}
	BM25(d, q) = \sum_{i = 1}^M \dfrac{IDF(t_i) \cdot TF(t_i, d)       \cdot (k_1+1)}{TF(t_i, d) + k_1 \cdot \left(1-b+b \cdot           \dfrac{len(d)}{avgdl}\right)}
	\label{eq:BM25 calculation}
        \end{equation}        
        where $k_1$, $b$ are free hyperparameters. $TF$ denotes term frequency and $IDF$ signifies inverse document frequency. The term $avgdl$ is the represents the average document length across the dataset.

    \item[-] \textbf{QLD}~\cite{zhai2008statistical} is an efficient probabilistic statistical model, assesses relevance scores by evaluating the likelihood of query generation.
    The computation of the QLD score is outlined as follows:

   \begin{equation}
    	\log p(q|d) = \sum_{i: c(q_i; d)>0} \log \dfrac{p_s(q_i|d)}{\alpha_d p(q_i|\mathcal{C})} + n \log \alpha_d +\sum_i \log p(q_i|\mathcal{C})
    	\label{eq:language model calculation}
    \end{equation}

    For more information, please refer to Zhai et al.'s work\cite{zhai2008statistical}.

    \item[-] \textbf{BM25\_ngram} is a modified version of BM25 in order to better determine relevance through lexical matching. Given the abundance of uncommon specialized terms in legal case documents, which hold unique meanings in specific contexts, specific combinations of terms can offer fresh insights into relevance identification. Therefore, we implemented Bm25\_ngram by adapting the ngram\_range parameter of the TfidfVectorizer. The ngram\_range parameter specifies the lower and upper boundaries for the range of n-values corresponding to different n-grams to be extracted. 
\end{itemize}

\begin{figure*}[t]
\centering
\includegraphics[width=0.8\textwidth]{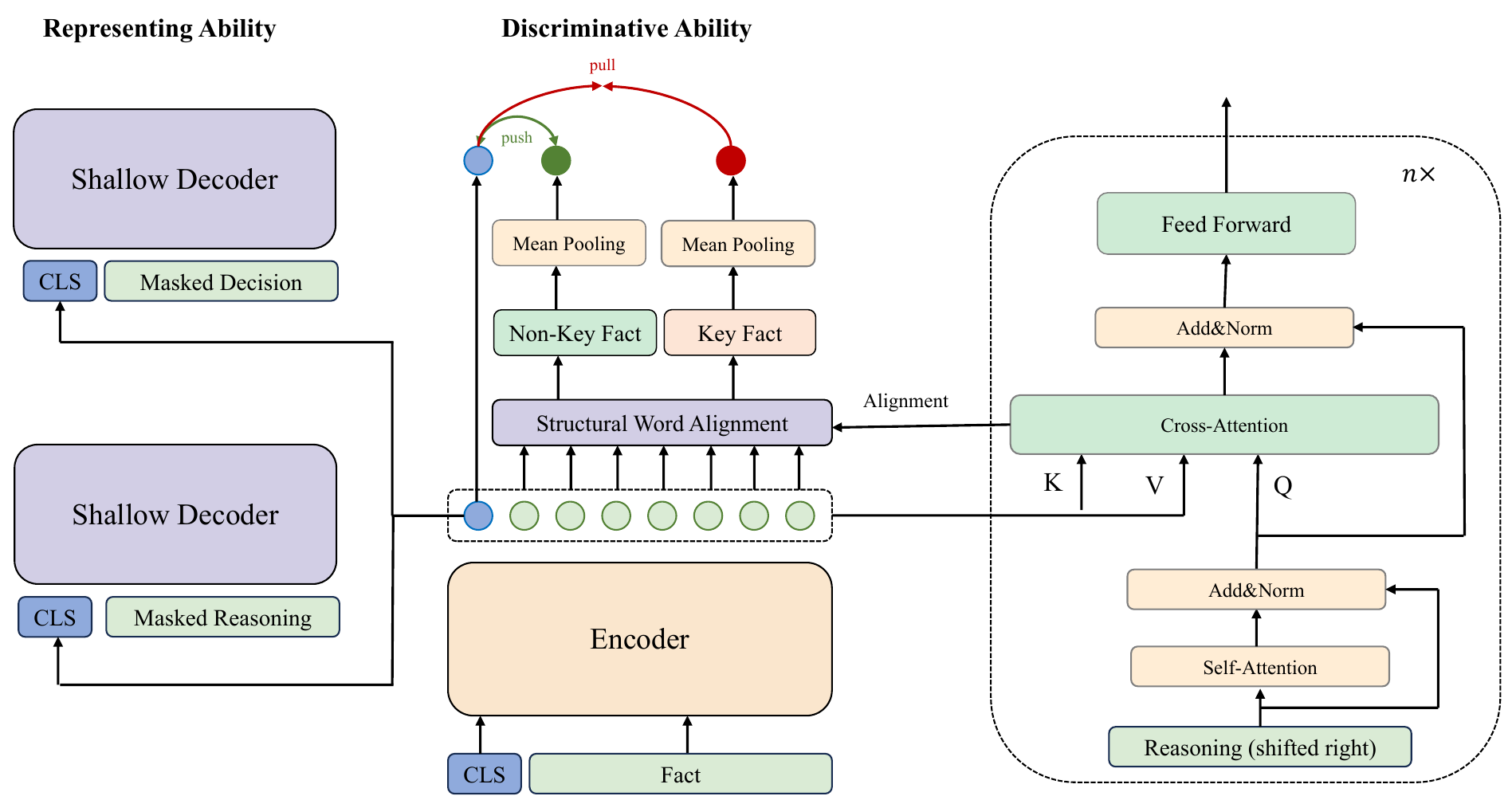}

\caption{Pre-training designs of DELTA. }
\label{model}
\end{figure*}

\subsubsection{Semantic Retrieval Models}
Semantic retrieval models can effectively avoid the problem of lexical mismatch and have been widely used in legal retrieval. However, pre-trained language models often perform unsatisfactory due to the limited input length and the difficulty of effectively understanding legal structures. Recently, a series of work has achieved state-of-the-art results by designing specific pre-training objectives for legal case retrieval. In this section, we implement SAILER and optimize it for better identification of legal case relevance.

\begin{itemize}[leftmargin=*]
    \item \textbf{SAILER}~\cite{li2023sailer} is a structure-aware pre-trained model. It fully utilizes the structure of legal documents to construct information bottlenecks and achieves state-of-the-art results on legal case retrieval tasks. We continued to fine-tune SAILER with the training sets of COLIEE2023 and COLIEE2022.
    \item \textbf{DELTA}~\cite{li2024delta} is an improved version of SAILER, which enhances the understanding of key facts in the legal cases and improves the discriminatory ability. To be specific, DELTA introduces a deep decoder which implements the translation of Fact section to Reasoning section. Afterwards, the word alignment mechanism is employed to determine key facts. Following this, the representation of the case in the vector space is pulled closer to the key facts and pushed away from the non-key facts. The framework of DELTA is shown as Figure~\ref{model}.
\end{itemize}

\begin{table*}[t]
\centering
\scriptsize
\caption{Features employed in our learning-to-rank approach for COLIEE2024 Task1. The placeholder contains ``FRAGMENT\_SUPPRESSED", ``REFERENCE\_SUPPRESSED",  ``CITATION\_SUPPRESSED".}
\begin{tabular}{cll}
\hline
\multicolumn{1}{l}{Feature ID} & Feature Name      & Description                                  \\ \hline
1                              & query\_length     & Length of the query                          \\
2                              & candidate\_length & Length of the candidate paragraph            \\
3                              & query\_ref\_num              & Number of placeholders in the query case           \\
4                              & doc\_ref\_num              & Number of placeholders in the candidate case             \\

5                              & BM25              & Query-candidate scores with BM25 (k\_1 = 3.0 , b = 1.0)             \\
6                              & BM25\_rank              & Rank of documents in the search list of the query by BM25 score            \\
7                              & QLD               & Query-candidate scores with QLD              \\
8                              & QLD\_rank              & Rank of documents in the search list of the query by QLD score            \\

9                              & BM25\_ngram               & Query-candidate scores with BM25\_ngram             \\
10                              & BM25\_ngram\_rank              & Rank of documents in the search list of the query by BM25\_ngram score            \\
11                              & SAILER        & Inner product of query and candidate vectors generated by SAILER       \\
12                              & SAILER\_rank              & Rank of documents in the search list of the query by SAILER score            \\
13                              & DELTA        & Inner product of query and candidate vectors generated by DELTA       \\
14                              & DELTA \_rank              & Rank of documents in the search list of the query by DELTA score            \\
 \hline
\end{tabular}
\label{feauture}
\end{table*}

\subsubsection{Learning to Rank}

Following previous work~\cite{yang2022thuir,li2023towards,chen2023thuir,tu2023thuir,han2023thuir_ss}, lWe utilize Lightgbm to integrate all feature scores.
Table~\ref{feauture} shows the details of all the features. A total of 14 features were used to integrate the final score. For optimizing ranking, we employ the Normalized Discounted Cumulative Gain (NDCG) as our objective. The model demonstrating the highest performance on the validation set is selected for subsequent testing.

\subsubsection{Post-processing}
Finally, we post-processed the ranking scores from the relevance perspective to remove irrelevant documents. Apart from Filtering by trial date, Filtering query cases and Dynamic cut-off proposed in previous li et al. work~\cite{li2023thuircoliee}, we add Filtering duplicate cases as a post-processing strategy. The specific details are as follows:

\begin{itemize}[leftmargin=*]
    \item \textbf{Filtering by trial date.} Considering that a query case typically cites cases preceding its trial date, it is logical to filter the candidate set based on this criterion. By extracting all dates mentioned within each case, we determine the latest date as the trial date, thereby minimizing erroneous exclusions. In instances where dates cannot be extracted from query cases, we retain all cases in the candidate set.
    \item  \textbf{Filtering query cases.} We find that query cases hardly become noticed case for other queries. Therefore we remove all query cases from the search results.
    \item  \textbf{Filtering duplicate cases.}  We find that all the noticed cases are not repeating in the COLIEE2021, COLIEE2022 and COLIEE2023 query cases respectively, indicating that deleting duplicate cases might be effective. Kim et al.~\cite{kim2024legal} also used removing repeating cases in the previous retrieval task, utilizing maximum duplicate cases as the hyper-parameter. By noticing that removing duplicate cases may delete all the candidate cases for some query cases, we define $t$ as the maximum numbers of duplicate cases and then supplement $s$ cases with higher score for those query cases without candidate case. Grid search in the validation set is utilized to find optimal $t$ and $s$.    
    \item \textbf{Dynamic cut-off} To accommodate the variability in the number of supporting cases associated with different query cases, we implement a dynamic-cutoff mechanism for each query case. This involves defining three hyperparameters: $h$, $l$, and $p$, respectively.  Here, $h$ represents the maximum, and $l$ the minimum number of supporting cases to be retrieved per query case. Additionally, if the highest score achieved by supporting cases for a specific query case is denoted as $S$, then only those supporting cases scoring above $p \times S$ are selected. A grid search technique is employed to ascertain the optimal values for these hyperparameters $h$,$l$ and $p$.
\end{itemize}

\subsection{Task3.The Statute Law Retrieval Task}
In this section, we follow the framework of Task 1 to implement Task 3. Specifically, we design heuristic pre-processing and post-processing strategies and implement advanced retrievers and rankers. Finally, we use learning to rank to integrate all scores.

\subsubsection{Pre-processing}
In Task 3, we primarily pre-process the retrieval pool, i.e., the legal articles. Specifically, we started by removing the lead-in information from the Civil Code. For example: ``Part I General Provisions'',
``Chapter I Common Provisions''. We consider that this information does not contribute to the relevance judgment.
Subsequently, we deleted all explanatory descriptions in brackets, such as (Standards for Construction). We consider that these are too general and do not facilitate the differentiation of legal articles.
Finally, we obtain a mapping of article IDs and specific content to form the retrieval set.

\subsubsection{Retriever}
We implemented the following retriever to get the most relevant legal articles from the full set:

\begin{itemize}[leftmargin=*]
    \item \textbf{BM25}~\cite{robertson2009probabilistic} is a robust lexical matching method. In Task 3, we set $k_1$ to 0.99 and $b$ to 0.75.

    \item \textbf{QLD}~\cite{zhai2008statistical} is another effective probabilistic lexical model. The detailed description can be found in section 4.1.

\end{itemize}

\subsubsection{Reranker}
After getting the retrieved $top200$ relevant legal articles, we use reranker to further rank them. The detailed model is as follows:

\begin{itemize}[leftmargin=*]
    \item \textbf{BERT}~\cite{devlin2018bert} is the classic pre-trained language model, which employs a multi-layer bidirectional Transformer encoder architecture, BERT leverages both the Masked Language Model (MLM) and Next Sentence Prediction (NSP) as its pre-training tasks.
    \item \textbf{RoBERTa}~\cite{liu2019roberta}  represents an advancement over BERT, utilizing a more extensive dataset for pre-training. Unlike BERT, RoBERTa is exclusively pre-trained using the Masked Language Model (MLM) task.
    \item \textbf~\cite{chalkidis2020legal} has been pre-trained on an extensive English legal database and has demonstrated state-of-the-art performance across a variety of legal tasks.
    \item \textbf{monoT5}~\cite{nogueira2020document} adopts an encoder-decoder architecture. It operates by generating a ``true'' or ``false'' token, reflecting the relevance between queries and candidates. The model then considers the probability of generating ``true'' as the ultimate relevance score.

\end{itemize}

For BERT, RoBERTa, and LEGALBERT, we train them with the cross-encoder architecture. Specifically, the query and legal articles are spliced together and fed into the encoder, and the vector of $[CLS]$ token is passed through the MLP layer to get the final score. The loss function for training is as follows:
\begin{equation}\label{eqn-1} 
  L(q,d^+,d^-_{1},...,d^-_{n}) =
-\log_{}{    \frac{exp(s(q,d^+))}{exp(s(q,d^+))+\sum_{j=1}^nexp(s(q,d^-_j))}}
\end{equation}
where $d^{+}$ and $d^{-}$ are relevant and negative articles. We employ irrelevant articles from the $top200$ articles retrieved by BM25 as negative examples.
For monoT5, we trained three versions of monoT5\_base, monoT5\_large, and monoT5\_3B.

\subsubsection{Learning to Rank}
Similar to Task 1, we integrate all the features using Lightgbm. The features utilized in Task 3 are displayed in Table~\ref{task3}. A total of 9 features were employed to integrate the final score. We adopt $Precision@1$ as the optimization objective and select the best model based on performance on the validation set for testing purposes.

\begin{table*}[t]
\centering
\caption{Features that we used for learning to rank in COLIEE2024 Task 3.}
\begin{tabular}{cll}
\hline
\multicolumn{1}{l}{Feature ID} & Feature Name      & Description                                  \\ \hline
1                              & query\_length     & Length of the query                          \\
2                              & article\_length & Length of the candidate article            \\
3                              & BM25              & Query-article scores with BM25             \\
4                              & QLD               & Query-article scores with QLD              \\
5                              & BERT        & Query-article scores with BERT       \\
6                              & RoBERTa     & Query-article scores with RoBERTa   \\
7                              & LEGALBERT    & Query-article scores with LEGAL-BERT-base   \\
8                              & monoT5\_large   & Query-article scores with monoT5\_large \\
9                              & monoT5\_3B         & Query-article scores with monoT5\_3B        \\ \hline
\end{tabular}
\label{task3}
\end{table*}

\subsubsection{Post-processing}
Finally we performed the heuristic post-processing on the ranking scores. Upon analysis, it was observed that the majority of queries are associated with no more than two relevant legal articles.
Therefore, we define the maximum score for one query to be $S$. Only articles that exceed the $S \times p$ score are considered relevant.
The hyperparameter $p$ is finely tuned to maintain consistency in the proportion of queries with two relevant laws across both the training and validation sets.

\begin{table}[t]
\centering
\caption{Performance and optimal hyperparameter on COLIEE2024 validation set.}
\begin{tabular}{lllllllll}
\hline
Model     & F1 score & Precision & Recall & p    & h & l & t & s \\ \hline
TQM\_run1 & 0.3824   & 0.3708    & 0.3046 & 0.7  & 5 & 4 & 1 & 2 \\
TQM\_run2 & 0.4294   & 0.4064    & 0.4552 & 0.3  & 7 & 4 & 1 & 2 \\
TQM\_run3 & 0.4592   & 0.4530    & 0.4656 & 0.46 & 7 & 1 & 1 & 2 \\ \hline
\end{tabular}
\label{task1_valid}
\end{table}

\begin{table}[ht]
\centering
\caption{Results on the official test of COLIEE2024 Task 1. Best results are marked bold.}
\small
\begin{tabular}{lcccc}
\hline
Team    & Submission                      & \multicolumn{1}{c}{F1} & \multicolumn{1}{c}{Precision} & \multicolumn{1}{c}{Recall} \\ \hline
TQM     & task1\_test\_answer\_2024\_run1 & \textbf{0.4432}        & \textbf{0.5057}               & 0.3944            \\
TQM     & task1\_test\_answer\_2024\_run3 & 0.4342                 & 0.5082                        & 0.3790                     \\
UMNLP   & task1\_umnlp\_run1              & 0.4134                 & 0.4000                        & 0.4277                     \\
UMNLP   & task1\_umnlp\_run2              & 0.4097                 & 0.3755                        & 0.4507                     \\
UMNLP   & task1\_umnlp\_runs\_combined    & 0.4046                 & 0.3597                        & 0.4622                     \\
YR      & task1\_yr\_run1                 & 0.3605                 & 0.3210                        & 0.4110                     \\
TQM     & task1\_test\_answer\_2024\_run2 & 0.3548                 & 0.4196                        & 0.3073                     \\
YR      & task1\_yr\_run2                 & 0.3483                 & 0.3245                        & 0.3758                     \\
YR      & task1\_yr\_run3                 & 0.3417                 & 0.3184                        & 0.3688                     \\
JNLP    & 64b7b-07f39                     & 0.3246                 & 0.3110                        & 0.3393                     \\
JNLP    & 07f39                           & 0.3222                 & 0.3347                        & 0.3105                     \\
JNLP    & 64b7b-48fe5                     & 0.3103                 & 0.3017                        & 0.3195                     \\
WJY     & submit\_1                       & 0.3032                 & 0.2700                        & 0.3457                     \\
BM24    & task1\_test\_result             & 0.1878                 & 0.1495                        & 0.2522                     \\
CAPTAIN & captain\_mstr                   & 0.1688                 & 0.1793                        & 0.1594                     \\
CAPTAIN & captain\_ft5                    & 0.1574                 & 0.1586                        & 0.1562                     \\
NOWJ    & nowjtask1run2                   & 0.1313                 & 0.0895                        & 0.2465                     \\
NOWJ    & nowjtask1run3                   & 0.1306                 & 0.0957                        & 0.2055                     \\
NOWJ    & nowjtask1run1                   & 0.1224                 & 0.0813                        & 0.2478                     \\
WJY     & submit\_3                       & 0.1179                 & 0.0870                        & 0.1831                     \\
WJY     & submit\_2                       & 0.1174                 & 0.0824                        & 0.2042                     \\
MIG     & test1\_ans                      & 0.0508                 & 0.0516                        & 0.0499                     \\
UBCS    & run3                            & 0.0276                 & 0.0140                        & \textbf{0.7196}                     \\
UBCS    & run2                            & 0.0275                 & 0.0140                        & 0.7177                     \\
UBCS    & run1                            & 0.0272                 & 0.0139                        & 0.7100                     \\
CAPTAIN & captain\_bm25                   & 0.0019                 & 0.0019                        & 0.0019                     \\ \hline
\end{tabular}
\label{task1_test}
\end{table}

\section{EXPERIMENT RESULT}
In this section, we present the results of our experiments and the corresponding analysis.
\subsection{Task1.The Case Law Retrieval Task}

\subsubsection{Submissions}
For COLIEE2024 Task 1, we submitted 3 runs with the following details
\begin{itemize}[leftmargin=*]
    \item \textbf{task1\_test\_answer\_2024\_run1}: We implemented the lexical matching model QLD and searched for the best parameters $t,s,h,l,p$ on the validation set based on the QLD scores in the post-processing stage and applied them to the test set.
    \item \textbf{task1\_test\_answer\_2024\_run2}: The improved lexical matching model BM25\_ngram was implemented, and an optimal set of parameters $t,s,h,l,p$ was identified through a search on the validation set, guided by the BM25\_ngram scores during the post-processing stage. These parameters were subsequently applied to the test set.
    \item \textbf{task1\_test\_answer\_2024\_run3}: The lightgbm integrates all the features to get the final score, after which the best post-processing parameters are obtained based on this score and applied to the test set.

\end{itemize}

\begin{table}[ht]
\centering
\caption{The performance of various model on COLIEE2024 task3 validation set. Best results are marked bold.}
\begin{tabular}{lccc}
\hline
Model         & F2              & Precision       & Recall          \\ \hline
BM25          & 0.5267          & 0.6039          & 0.5181          \\
QLD           & 0.3888          & 0.4257          & 0.3844          \\
BERT          & 0.6698          & 0.7524          & 0.6600          \\
RoBERTa       & 0.6637          & 0.7524          & 0.6534          \\
LEGALBERT     & 0.6929          & 0.7920          & 0.6815          \\
monoT5\_base  & 0.6951          & 0.7821          & 0.6848          \\
monoT5\_large & 0.7072          & 0.8019          & 0.6963          \\
monoT5\_3B    & \textbf{0.7171} & \textbf{0.8118} & \textbf{0.7062} \\ \hline
\end{tabular}
\label{task3_valid}
\end{table}

\begin{table*}[t]
\centering
\caption{Results on the official test of COLIEE2024 Task 3.Best results are marked bold. * indicates runs that use LLMs with undiscolsed training data. · indicates runs that use LLMs with discolsed training data. \# is runs without LLM.}
\begin{tabular}{lccccccc}
\hline
Submission\_id               & F2              & Precision       & Recall          & MAP             & R\_5            & R\_10           & R\_30           \\ \hline
JNLP.constr-join*            & \textbf{0.7408} & 0.6502          & \textbf{0.7982} & 0.8010          & \textbf{0.8769} & \textbf{0.9154} & 0.9462          \\
CAPTAIN.bjpAllMonoT5·        & 0.7335          & 0.6713          & 0.7752          & \textbf{0.8149} & 0.8615          & 0.9308          & 0.9538          \\
TQM-run1\#                   & 0.7171          & \textbf{0.7202} & 0.7339          & 0.7899          & 0.8308          & 0.9000          & \textbf{0.9615} \\
CAPTAIN.bjpAllMonoP·         & 0.7171          & 0.6743          & 0.7477          & 0.7731          & 0.8538          & 0.9308          & 0.9538          \\
CAPTAIN.bjpAll\#             & 0.7135          & 0.6227          & 0.7844          & 0.8149          & 0.8615          & 0.9308          & 0.9538          \\
JNLP.Mistral*                & 0.7123          & 0.6682          & 0.7477          & 0.7434          & 0.8308          & 0.9154          & 0.9538          \\
NOWJ-25mulreftask-ensemble\# & 0.7081          & 0.6334          & 0.7661          & 0.7562          & 0.8231          & 0.8769          & 0.9077          \\
AMHR02·                      & 0.6876          & 0.5972          & 0.7569          & 0.7405          & 0.7846          & 0.8308          & 0.8462          \\
AMHR03·                      & 0.6825          & 0.6456          & 0.7202          & 0.7405          & 0.7846          & 0.8308          & 0.8462          \\
AMHR01·                      & 0.6749          & 0.5734          & 0.7569          & 0.7405          & 0.7846          & 0.8308          & 0.8462          \\
NOWJ-25multask-ensemble\#    & 0.6654          & 0.5934          & 0.7431          & 0.7180          & 0.7231          & 0.8077          & 0.8692          \\
NOWJ-25mulref-ensemble\#     & 0.6649          & 0.5916          & 0.7202          & 0.7315          & 0.8154          & 0.8462          & 0.8923          \\
TQM-run2\#                   & 0.6621          & 0.5734          & 0.7110          & 0.7082          & 0.7769          & 0.8077          & 0.8077          \\
JNLP.RankLLaMA*              & 0.6555          & 0.6606          & 0.6651          & 0.7400          & 0.8385          & 0.9154          & 0.9538          \\
UA-mp\_net\#                 & 0.6409          & 0.4908          & 0.7385          & 0.7127          & 0.8000          & 0.8538          & 0.9000          \\
UA-anglE\#                   & 0.6399          & 0.4679          & 0.7477          & 0.6935          & 0.7538          & 0.8077          & 0.8769          \\
TQM-run3\#                   & 0.6330          & 0.5963          & 0.6606          & 0.7492          & 0.8154          & 0.8692          & 0.9308          \\
BM24-1*                      & 0.4945          & 0.2590          & 0.7294          & -               & -               & -               & -               \\
MIG2\#                       & 0.1665          & 0.1604          & 0.1881          & 0.2125          & 0.2615          & 0.2923          & 0.3769          \\
MIG1\#                       & 0.1637          & 0.1187          & 0.2064          & 0.2049          & 0.2385          & 0.2923          & 0.3846          \\
MIG3\#                       & 0.1629          & 0.1631          & 0.1789          & 0.2049          & 0.2385          & 0.2923          & 0.3846          \\
PSI01                        & 0.0785          & 0.0826          & 0.0780          & 0.2312          & 0.3692          & 0.4769          & 0.6308          \\ \hline
\end{tabular}
\label{task3_test}
\end{table*}

\subsubsection{Results}

Table~\ref{task1_valid} shows the effectiveness and optimal parameters of submission runs on the validation set. Table~\ref{task1_test} shows the final official evaluation results. From the experimental results, we can draw the following conclusions:

\begin{itemize}[leftmargin=*]
    \item From the results of the validation set, the lexical matching model Bm25\_ngram achieved competitive results. Learning to rank effectively combines the perspectives of lexical matching model and semantic retrieval model to achieve the best results.
    \item However, the official test results showed different performance. BM25\_ngram had the worst results and QLD achieved the best performance. We speculate this is due to the bias in the distribution of terms on the test sets of COLIEE2023 and COLIEE2024. Since the distribution of the BM25\_ngram scores is different on the two datasets, it results slightly lower performance of learning to rank than the single model.
    \item  Overall, our approach achieves championship in the legal case retrieval task and shows sufficient robustness, which is crucial in legal scenarios where large-scale annotation data is lacking.

\end{itemize}

\subsection{Task3.The Statute Law Retrieval Task}
\subsubsection{Submissions}
In Task 3, we submit 3 runs as follows:

\begin{itemize}[leftmargin=*]
    \item \textbf{TQM\_run1}: We fine-tuned monoT5\_3B using the training data and performed post-processing.
    \item \textbf{TQM\_run2}: Lightgbm was employed to integrate all features and use Precision@1 as the optimization objective.
    \item \textbf{TQM\_run3}: Lightgbm was employed to integrate all features and use Precision@2 as the optimization objective.
\end{itemize}

\subsubsection{Results}
Table~\ref{task3_valid} shows the performance of various models on the validation set. Table~\ref{task3_test} shows the official evaluation results.
We derive the fol- lowing observations from the experiment results.
\begin{itemize}[leftmargin=*]
    \item From the Table~\ref{task3_valid} , it can be observed that Ranker performs better than the Retriever. The best single model result was achieved by mono\_T5.
    \item However, the performance drops significantly after learning to rank on the test set. We think this is due to overfitting caused by too little training data. How to effectively integrate each feature deserves further research.
    \item  Overall, our submission had the best performance among all the runs without LLMs, and ranked third among all the submissions. This suggests that LLMs can be effective in enhancing the understanding of the law thus improving the performance.

\end{itemize}

\section{Conclusion}
This paper presents TQM Team’s approaches to the legal case retrieval task in the COLIEE 2024 competition. We try to enhance the understanding of the model for case relevance from multiple perspectives and achieve some progress. We obtained the best performance in Task 1 among all submissions, and the third place in Task 3. In the future we will continue to explore infusing legal knowledge into the model to better understand case relevance.

%
%
%

\bibliographystyle{splncs04}
\bibliography{sample.bib}




\end{document}